\documentclass[11pt,sort&compress]{elsarticle}
\usepackage{amsmath,amssymb,epsfig,multirow,xcolor}
\usepackage[normalem]{ulem}

\newcommand{\txf}{\textbf}

\newcommand{\veps}{\varepsilon}

\begin{document}

\title{A regularized Matched Interface and Boundary Method (MIB) for Solving Polarizable Multipole Poisson-Boltzmann model
}
\author[smu]{Xin Yang}
\ead{xiny@smu.edu}

\author[ua]{Shan Zhao}
\ead{szhao@ua.edu}

\author[smu]{Weihua~Geng\corref{cor1}}
\ead{wgeng@smu.edu}

\cortext[cor1]{Corresponding author}

\address[smu]{Department of Mathematics, Southern Methodist University, Dallas, TX 75275 USA}
\address[ua]{Department of Mathematics, University of Alabama, Tuscaloosa, AL 35487 USA}

\begin{abstract}
\noindent
To accurately model the electron density and polarization, 
a polarizable multipole (PM) model using the AMOEBA force field has been introduced \cite{Ren:2003, Shi:2013} recently. 
In the AMOEBA force field, 
the traditional point atomic representation is updated with permanent multipoles including additional dipoles and quadrupoles 
at atom centers in terms of derivatives of delta functions.  
Meanwhile, the polarization of the solute 
is considered by the introduction of induced dipoles. 
The AMOEBA forcefield thus shows significantly better 
agreement with experimental and high-level {\it ab initio} results. 
Moreover, the AMOEBA force field keeps the simple atomic structure, 
so that it  can conviniently replace the traditional partial charge model. 

In this paper, we address the numerical challenges 
associated with the Polarizable Multipole Poisson--Boltzamnnn (PM-PB) model, 
which couples the AMOEBA force field with a linear Poisson-Boltzmann equation  
for implicit solvent and polarization modeling. 
To solve the PM-PB model, 
we designed a regularized Matched Interface and Boundary (MIB) method 
to analytically regularizes  the singular source term in the PMPB model 
while maintains 2nd order accuracy by rigorously treating the interface conditions. 
The accuracy of the method is validated on Kirkwood sphere with available analytical solutions 
and on proteins whose charge distribution are assigned using AMOEBA force field.

\end{abstract}


\maketitle

%

\section{Introduction}
%

The Poisson-Boltzmann (PB) model~\cite{Warwicker:1982, Sharp:1990a,  Honig:1995} 
is a widely used implicit solvent model for studying electrostatic interaction 
between a solute, such as  protein, DNA, and RNA, and its surrounding solvent environment. 
Such an electrostatic analysis is indispensable for understanding various solvated biological processes at atomic level, 
including DNA recognition, transcription, translation, protein folding, protein ligand binding, etc. 
In spite of the great success that the classical PB model has achieved 
for electrostatic and solvation analysis \cite{Baker:2005,Lu:2008}, 
numerous improved PB models have been developed in the literature 
to reduce the modeling errors
by capturing as many atomic details as possible into the continuum electrostatics,
including 
differential geometry based multiscale model \cite{Wei:2010}, 
variational implicit solvent model \cite{Zhou:2014}, 
size-modified PB model \cite{Wang:2014}, 
nonlocal PB model \cite{Xie:2016}, 
heterogeneous dielectric model \cite{Li:2014,Hazra:2019}, to name just a few. 

{This manuscript  focuses on} reducing the modeling errors
due to the charge density representation for the source term of the PB equation,
which in the most accurate form should be an electron density distribution, 
obtained through expensive quantum mechanical (QM) calculations \cite{Tannor:1994,Mei:2006}. 
The most popular PB source in the literature is the partial charge model \cite{Sharp:1990a,  Honig:1995}, 
in which point charges located at atomic centers in terms of Dirac delta functions 
are summed  to approximate 
the QM charge density \cite{Sigfridsson:1998}. 
Benchmarked with experimental data, the partial charges of a protein can be directly generated through force field definitions  \cite{Baker:2005}. 
However, such discrete charge representation is known to be  an important source of the PB modeling errors \cite{Stone:2013} 
and is unable to capture the important polarization \cite{Sagui:2004}, 
i.e., the redistribution of the electron density in the presence of an external electric field. 

For improved accuracy of modeling electron density and polarization while keeping the atomic representation, 
a polarizable multipole (PM) model using the AMOEBA (atomic multipole optimized energetics for biomolecular applications) force field has been introduced \cite{Ren:2003, Shi:2013}. 
In the AMOEBA, permanent multipoles including dipoles and quadrupoles are assumed at atom centers in terms of derivatives of delta functions, 
and polarization of the solute is programmed for calculating induced dipoles. Thus, this model shows significantly better 
agreement with experimental and high-level {\it ab initio} results in cluster structures, 
energetics, bulk thermodynamic, structural measures for water \cite{Ren:2003}, organic molecule \cite{Ren:2011}, protein \cite{Shi:2013}, etc.  
Moreover, the AMOEBA force field keeps the simple atomic structure, so that it  can seamlessly replace the traditional partial charge model as the PB source. 

This manuscript target the numerical solution to the PMPB model, 
which couples the AMOEBA force field with a linearized PB equation for implicit solvent modeling. 
Various numerical challenges are faced in this development.  
Analytically, 
the multipole charges in terms of Dirac delta functions 
and their derivatives introduce singular sources for the PB equation, 
which poses a significant difficulty for the theoretical analysis \cite{Chen:2007}. 
Numerically, the commonly used charge distribution method \cite{Nicholls:1991}, 
which approximates singular point charges by finite values on surrounding grid nodes, 
leads to significant accuracy reduction. 
The work involved in this manuscript aims to address these challenges. 
Due to the numerical difficulties in solving the PB model under the AMOEBA framework, to our best knowledge, there are only two previous work from Schnieders et al. \cite{Schnieders:2007} and Cooper \cite{Cooper:2019} addressing the PMPB model. The former is under  3-d finite difference framework using B-spline interpolation to address the charge distribution while the latter is under the boundary integral framework on 2-d molecular surface only. 
The current work utilizes the charge regularization schemes \cite{Geng:2017} and treatment of complex geometry \cite{Yu:2007}, resulting a second order convergence scheme on 3-d volumetric meshes. 

The rest of the manuscript is organized in the following. In the next section, 
theory and algorithm are introduced,  
including the PB model, 
the AMOEBA model, 
and the PMPB model, 
followed by MIB method and the algorithm for solving the PMPB model using MIB. 
Polarization and its numerical treatment will be addressed too. 
Numerical results will follow with simple cases that have analytical solutions 
and real-world protein cases that are tested with convergence patterns. 
The manuscript ends with a section of concluding remarks. 

\section{Theory and Algorithm}

\subsection{The Poisson-Boltzmann Model}
Consider the 
linearized Poisson-Boltzmann (PB) equation in its most commonly used form \cite{Warwicker:1982, Sharp:1990a,  Honig:1995},
\begin{align}\label{LPBE}
&- \nabla \cdot (\veps \nabla \phi (\txf{r})) + \kappa^2 
\phi (\txf{r}) 
= \rho(\txf{r}), \quad \text{in} ~\,\Omega = \Omega^+ \cup \Omega^-,\\
& [\phi]:=\phi^+ - \phi^- = 0, \quad 
\left[\veps \frac{\partial \phi}{\partial  \nu} \right] 
:=\veps^+ \frac{\partial \phi^+}{\partial \nu} -\veps^- \frac{\partial \phi^-}{\partial \nu}
=0, \quad \mbox{on}~\, \Gamma, \quad \quad  \quad
\phi  =  g({\bf r}),  \quad  \mbox{on}~\,\partial \Omega \label{jump+BC}
\end{align}

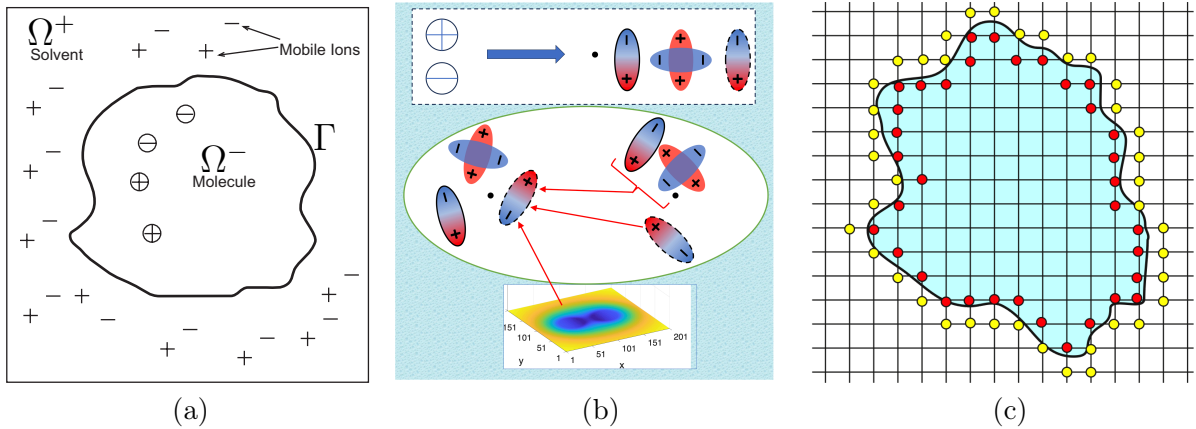
\begin{figure}[h!]
\begin{center}
        \includegraphics[width=1.9in]{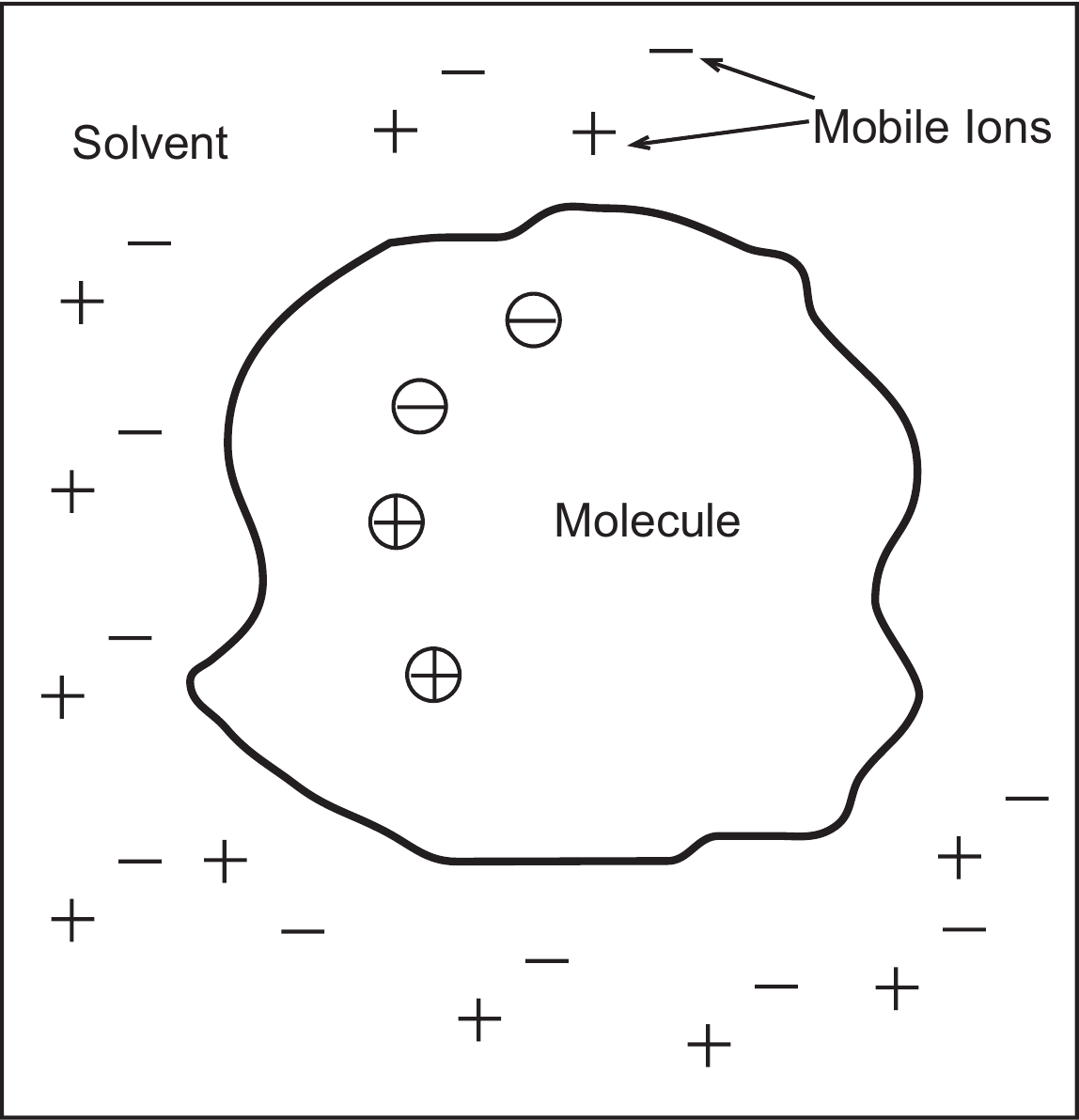}
	\setlength{\unitlength}{1cm}
	\begin{picture}(1,1)
	\put(   -4.8, 4.5){\Large $\Omega^+$}
	\put(  -1.0, 3.1){\Large $\Gamma$}
	\put( -2.5, 2.8){\Large $\Omega^-$}
	\end{picture}
	\hskip -30pt
	 \includegraphics[width=2.0in]{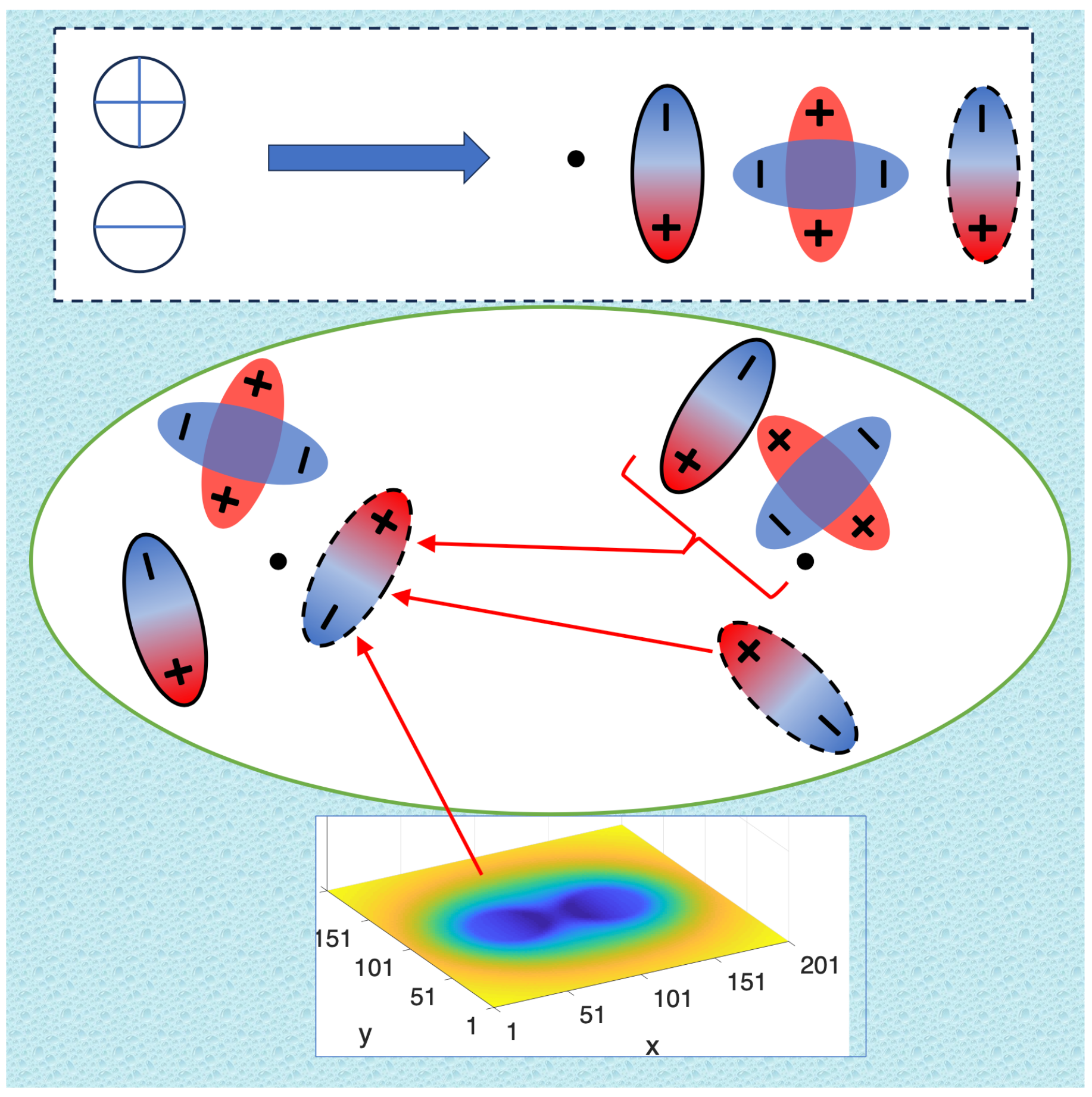}
	 \hskip 10pt
	  \includegraphics[width=2.0in]{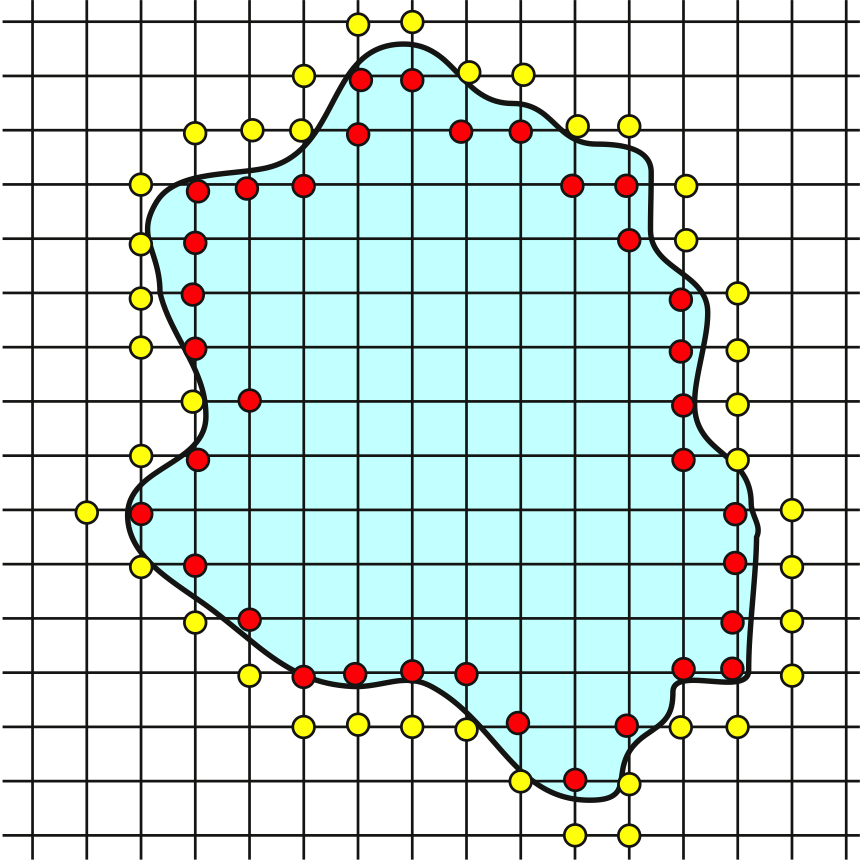}\\
	 (a) \hskip 1.9in (b) \hskip 1.9in (c) 
	 \caption{The illustration of PMPB Model: 
(a) Domains of the PB model with $\Omega^-$: a charged solute molecule, $\Omega^+$: solvent with mobile ions, and $\Gamma$: molecular surface; 
(b) An partial charge (a circled plus or minus sign) in the PB model is replaced by a multipole consisting of 
a monopole (a dot), 
a dipole (a solid-lined ellipse), 
and a quadrupole (two cross-intercepted ellipses), 
as well as an induced dipole (an dash-lined ellipse). 
For solvated molecules, 
besides direct induction and mutual induction, 
induced dipoles are also subject to polarization produced by the reaction field. 
In the AMOEBA force field, all multipoles are defined at atom centers. 
Here, components of the multipole are placed off-centers for illustration purpose.
(c) An 2-d illustration of the MIB schemes in which fictitious points (red/yellow)
near the molecular surface
are marked.
}
	\label{fig:domains}
\end{center}
\end{figure}
where $\phi$ is the electrostatic potential.  Figure~\ref{fig:domains}(a) illustrates the PB model in $\Omega=\Omega^- \cup \Omega^+ \subset \mathrm{R}^3$. 
Here $\Omega^-$ is the domain for the solute biomolecule and $\Omega^+$ is the domain for surrounding solvent containing dissolved mobile ions in terms of Boltzmann distribution. The dielectric interface $\Gamma$,  which separates $\Omega^-$ and $\Omega^+$, is usually defined as the solvent excluded surface (SES) \cite{Lee:1971,Richards:1977}. The Dirichlet boundary data $g({\bf r})$ in (\ref{jump+BC}) is calculated as a linear superposition 
of Coulomb's law or Screened Coulomb's potentail for all partial charges at each boundary node \cite{Holst:1994}.  In this manuscript, we address the PMPB model in which the partial charges in  Fig.~\ref{fig:domains}(a) are replaced by the polarizable atomic multipoles as illustrated in Fig.~\ref{fig:domains}(b). 
Note each atomic multipole consists a monopole, a dipole, and a quadrupole. 
To characterize the polarization, AMOEBA force field introduces induced dipole, whose polarization is produced by the permanent multipoles, other induced dipoles, and the reaction field as well. \underline{Note:} with the PMPB model, the $g({\bf r})$ needs to be modified accordingly using either single Debye-H\"uckel (SDH) or multipole Debye-H\"uckel (MDH) boundary conditions \cite{Schnieders:2007}. 

The three-dimensional numerical solution of the 
linearized PB (LPB) equation for biomolecules is known to suffer many difficulties, such as the following three items.\\
{\bf A}. The potential is singular at atom centers due to the charge sources 
\cite{Zhou:1996,Chern:2003,Chen:2007,Geng:2007,Holst:2012,Xie:2014,Geng:2017}.
The most commonly used source term assumes \emph{point charges}: 
$\rho(\txf{r}) = 4\pi C \sum_{n=1}^{N_c}q_n \delta (\txf{r}-\txf{r}_n)$, 
where $q_n$ is the $n$th partial charge located at ${\bf r}_n \in \Omega^-$
and $C$ is a constant depending on the units of the system. The source term $\rho(\txf{r})$ is more complicated in the 
PMPB model
in which point multipoles are involved.

\noindent
{\bf B}. The solute-solvent boundary is a geometrically complicated molecular surface  
\cite{Lee:1971,Richards:1977,Sanner:1996,Chen:2011a,Liu:2017}. 
The SES generation and the corresponding numerical mesh implementation 
are indispensable parts of the PB interface algorithms which will be provided in the next. 

\noindent
{\bf C}. An interface treatment is needed due to the loss of regularity in the potential $\phi$
\cite{Yu:2007,Chen:2011,Wang:2013,Deng:2015,Ahmed-Ullah:2020}. 
Across $\Gamma$, the dielectric constant or relative permittivity $\veps$ is piecewisely defined: $\veps =\veps^{\pm}$ for ${\bf r} \in \Omega^{\pm}$. For easily compare the numerical results, we follow the convention in \cite{Schnieders:2007} by letting $\veps^+ = 78.3$ and $\veps^-=1$. 
The Debye length ${\kappa}$ is also piecewise  with 
${\kappa} =0$ in $\Omega^-$ and 
${\kappa}=\bar{\kappa}:=8.486902807$\AA$^{-2} I_s$ in $\Omega^+$, where $\bar{\kappa}$ is
the modified Debye--H\"{u}ckel screening parameter and $I_s$ is the molar ionic strength.
Numerically, two interface jump conditions defined in (\ref{jump+BC}) with $\nu$ being the outward normal direction have to be satisfied in discretization so that accuracy reduction near $\Gamma$ can be avoided in the PB interface algorithms. 

\subsection{PB interface algorithms}\label{sec.interface}
The 3D elliptic interface problem associated with the PB equation (\ref{LPBE}) and (\ref{jump+BC}) is a well known challenge in the scientific computing literature with several obstacles. 
For a real protein, the solvent excluded surface (SES) has a complex shape.  Moreover, the SES is just Lipschitz continuous and may admit \emph{geometrical singularities}, such as cusps and tips \cite{Sanner:1996,Liu:2017}. Furthermore, although being continuous across the interface $\Gamma$ in (\ref{jump+BC}), {the potential and its flux become discontinuous} after the regularization \cite{Geng:2017,Lee:2020}. To deal with the Difficulties {\bf B} and {\bf C}, various PB interface algorithms have been developed. Below is a brief overview.  

 {\bf Body-fitted grid methods}. 
One elegant way to accommodate complex interfaces
and geometries in the PB simulations 
is fitting the grid to the material boundaries. 
Using unstructured grids, the finite element methods \cite{Holst:1995,Baker:2001,Chen:2007,Xie:2007,Xie:2014,Deng:2018}
and discontinuous Galerkin methods \cite{Bedin:2013,Yin:2014,Deng:2015}
are some of the most flexible methods for handling geometrically 
complex problems in electrostatic simulations. 
Based on evaluating integrals of irregular elements,
the finite-volume PB solution \cite{Baker:2001}
allows easy formulation for unstructured meshes too.
The use of fully automatic and adaptively refined multilevel 
tetrahedral meshes \cite{Cortis:1997,Holst:2001,Baker:2000}
in finite element simulations
greatly enhances the numerical accuracy near the material interfaces.
Similarly, graded and non-graded adaptive Cartesian grids 
based on octree structures \cite{Boschitsch:2011,Mirzadeh:2013} have been proposed
to resolve the material interface in finite differences.

{\bf Boundary element methods}. 
Based on Green's theorem, the linearized PB equation can be rewritten as a
boundary integral equation in terms of electrostatic potential and its flux. 
The boundary element methods 
\cite{Juffer:1991, Liang:1997,Boschitsch:2002, Lu:2006, Yokota:2011,Bajaj:2011,Geng:2013b,Geng:2015}
discretize the corresponding surface integrals over the 
2D triangularized dielectric interface, so that the Difficulties {\bf A} - {\bf C} can be analytically circumvented. 
The dense matrix computations can be accelerated by using fast algorithms,
such as fast multipole method (FMM)
\cite{Boschitsch:2002,Lu:2006,Yokota:2011,Bajaj:2011}
and treecode \cite{Geng:2013b}.

{\bf Finite difference interface methods}. 
The finite-difference method has been a mainstay for solving the PB equation 
in chemical and biological applications over the past few decades 
\cite{Im:1998,Rocchia:2001,Luo:2002}. 
To overcome the staircasing approximation to the arbitrarily shaped molecular surfaces, several  finite difference interface methods have been developed.
By assuming that the interface is aligned with a mesh line,
a jump condition capture scheme has been developed in \cite{Chern:2003}.
Based on a Cartesian grid, the immersed interface method (IIM) \cite{LeVeque:1994}
has been applied to solve the PB equation 
\cite{Qiao:2006,Wang:2009}, in which the jump conditions can be rigorously enforced based on Taylor expansions. 
The Matched Interface and Boundary (MIB) method to be adopted 
to solve the PMPB model falls into the category of finite difference interface methods, 
which will be explained briefly the next.

\subsection{Matched Interface and Boundary (MIB) Method} 
For the purpose of dealing with arbitrarily shaped  dielectric interfaces based
on a simple Cartesian grid,
a matched interface and boundary (MIB) PB solver
\cite{Zhou:2006a,Yu:2007,Geng:2007,Chen:2011,Geng:2017} 
has been developed through rigorous treatments of geometrical and charge singularities.
The MIB method is the known to be a second order convergent PB solver, 
which overcomes numerical difficulties brought by interface and singular sources.

We use the linearized PB equation as in Eq.~({\ref{ellipticeqn}) to explain 
the key ideas of the MIB method for solving
the elliptic interface problem with discontinuous coefficients although nonlinear PB equation can be efficiently solved too \cite{Chen:2011, Geng:2017}
\begin{equation}\label{ellipticeqn}
    -\nabla\cdot\left(\epsilon({\bf r})\nabla \phi({\bf r})\right)+\bar{\kappa}^2({\bf r}) \phi({\bf r})=\rho({\bf r}).
\end{equation}
As shown in Fig.~\ref{fig:domains}(a), the interface $\Gamma$  divides the whole domain into two
separated parts, $\Omega^-$ and $\Omega^+$. The jump conditions
across the interface are assumed to be
\begin{eqnarray}\label{jump1}
    [\phi]_\Gamma &=&  g_0({\bf r}), \\ \label{jump2}
    [\epsilon \phi_{\nu}]_\Gamma &=& g_1({\bf r}).
\end{eqnarray}
Note the physical meaning of $g_0$ and $g_1$ is the jump 
in electrostatic potential and flux density across the interface $\Gamma$, 
thus $g_0$ and $g_1$ are zeros in the physical background of electrostatic potential $\phi$. 
However, we keep the non-homogeneous form of $g_0$ and $g_1$ here to emphasize the capability of MIB method to treat the
non-homogeneous  interface jump conditions. 

Consider a uniform Cartesian grid partition of the domain $\Omega$ as shown in Fig.~\ref{fig:domains}(c),  
it is well known that the standard finite difference schemes lose
their designed convergence near the interface and the interface
jump conditions have to be used to restore the accuracy. 
To this end, all the grid points in $\Omega$ are classified into two types, the regular ones and the
irregular ones. An {\it irregular} grid point is defined as a node at which the standard finite difference scheme  
involves grid points across the interface, i.e., at least one of its four (2D) or six (3D) neighboring points is from
the other side of the interface as illustrated in Fig.~\ref{fig:domains}(c) for a 2D situation. 
The complement set to the set of irregular grid points defines the set of {\it regular} grid points. 
At a regular point, a centered difference discretization of Eq.~(\ref{ellipticeqn}) is carried out, 
which involves a grid node $(x_i,y_j,z_k)$ and its six neighboring points. 
At each irregular point, there are two values: the true value $\phi(x_i, y_j, z_k)$ and the fictitious value $f(x_i, y_j, z_k)$. 
The fictitious values can be considered as the extended value from one domain to the other, 
whose values are obtained by interpolation schemes involving both the differential equation and the interface jump conditions. 
In the MIB scheme, the finite difference approximations at irregular points will be modified by using
fictitious values from the other side of the interface. 
For example, if one needs to discretize $\frac{\partial^2 \phi}{\partial x^2} $ at $({x_{i+1},y_j,z_k})$ 
we have the
following modified finite difference approximation for the $x$ derivative
\begin{equation}\label{ModifiedFD}
\frac{\partial^2 \phi}{\partial x^2}|_{(x_{i+1},y_j,z_k)} \approx  \frac{1}{\Delta x^2} (f_{i,j,k} - 2\phi_{i+1,j,k}
+ \phi_{i+2,j,k}),
\end{equation}
where $f_{i,j,k}$ is a fictitious value defined at $(x_i,y_j,z_k)$.
The modified finite difference approximations at irregular points maintain the second order of accuracy, provided that the fictitious values are accurately estimated.

In the MIB scheme, 
by applying the interface jump conditions in Eqs.~(\ref{jump1}) and (\ref{jump2}), a pair of fictitious values $f_{i,j,k}$ at $(x_i,y_j,z_k)$ 
and $f_{i+1,j,k}$ at $(x_{i+1},y_j,z_k)$
will be represented as a linear combination of
function values on a set of neighboring nodes $\mathbb{S}_{i,j,k}$ and jump data
$(g_0,g_1)$. For example, 
\begin{equation}\label{FP}
f_{i,j,k} = \sum_{(x_I,y_J,z_K) \in \mathbb{S}_{i,j,k}} w_{I,J,K} \phi_{I,J,K} + w_0 g_0 + w_1 g_1.
\end{equation}
The major task of a particular MIB approximation is to determine the points set $\mathbb{S}_{i,j,k}$ 
and the representation weights $w_{I,J,K}$, $w_0$, and $w_1$ via discretizing Eqs.~(\ref{jump1}) and (\ref{jump2}).
We omit the details of finding fictitious values in MIB schemes, which can be found in \cite{Yu:2007}. 
Note for the situation very complicated geometry with sharp corners are encountered the size of $\mathbb{S}_{i,j,k}$, which is regularly 16, can be as big as 64 for 3-d case \cite{Yu:2007a, Yu:2007b}. 


Another issue is the regularization of the source singularity as seen on the right hand side of Eq.~(\ref{ellipticeqn}), where  
$\rho({\bf r})=4\pi C\sum\limits_{i=1}^{N_c} q_{i}\delta({\bf r}-{\bf r}_i)$. Our strategy is to use the Green's function based decomposition incorporated with the MIB schemes to treat the source singularities while maintain the 2nd order accuracy. Since this paper focuses on surface and interface treatment, we omit all the details for the  treatment of charge singularities, and interested readers can refer to \cite{Geng:2007, Geng:2017}. 
All MIBPB results reported in this work are based on the rMIB package developed in \cite{Geng:2017}. 

\subsection{The Polarizable Multipole Poisson-Boltzmann Model}
\subsubsection{Polarizable multipole (PM) sources of the AMOEBA}
We first establish the necessary notation for representing PM sources and the corresponding Green's function. 
Consider a protein with $N_c$ atoms, at the center of the $n${th} atom, i.e., ${\bf r}_n=(x_n,y_n,z_n)$,  
 the $n${th} permanent order 2 multipole ${\bf M}^n$ consists of 13 components:  
${\bf M}^n = [ q^n, d^n_x,  d^n_y,  d^n_z, Q^n_{xx}, Q^n_{xy}, \ldots, Q^n_{zz}]^T$, where $q, d_i, Q_{ij}$ for $i,j=x,y,z$ are the moments of the monopole, dipole, quadruple in suffix/Einstein notation.
Using this notation, the permanent charge at  ${\bf r}_n$ can be written as  \cite{Ren:2003, Shi:2013}
\begin{equation}\label{eq_qn}
	\rho^n({\bf r})=q^n\delta({\bf r}-{\bf r}_n)+d_i^n\partial_i\delta({\bf r}-{\bf r}_n)
	+Q^n_{ij}\partial_{ij}\delta({\bf r}-{\bf r}_n),
\end{equation}
A key idea of the singular charge regularization is to 
\underline{analytically capture the singularities} in $\rho$. 
For this purpose, the Coulomb potential $G^n$
governed by the Gauss's law 
\begin{equation}
- \Delta G^n = 4 \pi  \rho^n \label{eq_Gauss}
\end{equation}
in the free space is expressed in terms of the \emph{Green's function}
\begin{equation}\label{eq_Gn}
	G^n({\bf r})=\frac{1}{|{\bf r}-{\bf r}_n|}q^n + \frac{r_i - r_{n,i}}{|{\bf r}-{\bf r}_n|^3}d_i^n
+ \frac{(r_i - r_{n,i})(r_j - r_{n,j})}{2|{\bf r}-{\bf r}_n|^5} Q^n_{ij}.
\end{equation}
For all permanent multipoles ${\bf M} = [  {\bf M}^1, {\bf M}^2, \ldots, {\bf M}^{N_c}]^T$, the total Coulomb potential is \emph{additive} such as 
$G^{\bf M}({\bf r}) = \sum_{n=1}^{N_c} G^n({\bf r})$ by the superposition principle. 

\underline{Note:} In Eq.(~\ref{eq_Gauss}), we keep the term $4\pi$ with the charge $\rho^n$ at the right hand side, use vacuum dielectric constant 1, which thus is not shown, and leave out the vacuum permittivity $\epsilon_0$ (usually appeared as a denominator term for the right hand side in Eq.(~\ref{eq_Gauss}) and  in $G^n({\bf r})$).  By doing these, $G^n({\bf r})$ in Eq.~(\ref{eq_Gn}) has a very simple form, the units in Eq.(~\ref{eq_Gauss}) is $e_c/$\AA$^3$, and more importantly $G^n({\bf r})$ has the popular and physically interpretable unit of $e_c/$\AA ~for molecular simulation. By multiplying the coefficients 332.06364 (kcal/mol/$e_c$)(\AA$/e_c$) \cite{Holst:1994}, the potential will have the values using the classic unit of kcal/mol/$e_c$. 


\subsubsection{Polarization in the vacuum phase}
A simple model is proposed to describe polarization in the vacuum phase. 
Following the AMOEBA force field \cite{Ren:2003, Shi:2013}, only the dipole moments are polarizable, while quadruples are treated as non-polarizable for simplicity. The polarization is \emph{non-additive} even in vacuum thus the total polarization cannot be written as a sum of individual atomic polarizations. 
Based on the Green's function, we propose to calculate the induced dipole $\boldsymbol{\mu}^n$ at ${\bf r}_n$ as
\begin{equation}\label{eq_mu^n}
\boldsymbol{\mu}^n = \alpha^n {\bf E}^n = \alpha^n  \left( \sum_{m \ne n}  \nabla G^m({\bf r}_n) + 
\sum_{m \ne n}  {\bf T}_{nm} \boldsymbol{\mu}^m \right),
\end{equation}
where $\alpha^n$ is the isotropic atomic polarizability for the $n$th atom, and ${\bf E}^n$ is the electric field at ${\bf r}_n$. 
Here the gradient $ \nabla G^m({\bf r})$ is analytically available, and the tensor coefficient  ${\bf T}_{nm}$ is a 3-by-3 coefficient matrix given in the AMOEBA force field, subject to masking and Thole damping \cite{Schnieders:2007}. 

{\bf Two polarization parts}. The electric field ${\bf E}^n$ in Eq.~(\ref{eq_mu^n}) consists of two parts as shown in Fig. \ref{fig:domains}(b) (without the polarization induced from the reaction field illustrated at the bottom): 
(a) Polarization by other permanent multipoles, which is known as direct induction; 
(b) Polarization by other induced dipoles, which is known as mutual induction. 
The mutual polarization in Eq. (\ref{eq_mu^n}) is known as a \underline{self-consistent process}  \cite{Schnieders:2007}. Mathematically, the self-consistent process can be regarded as an iterative process: the $n${th} induced dipole $\boldsymbol{\mu}^n$ depends on all other induced dipoles $\boldsymbol{\mu}^m$, while the new value of $\boldsymbol{\mu}^n$ calculated by Eq. (\ref{eq_mu^n}) will in turn affect other induced dipoles $\boldsymbol{\mu}^m$. For electrostatic analysis, the equilibrium state of the self-consistent process needs to be calculated. 

{\bf Computation}. 
Equation~(\ref{eq_mu^n}) can be written as $M \boldsymbol{\mu}^\text{(V)} = {\bf g}$ where the superscript $\text{(V)}$ stands for vacuum with $\boldsymbol{\mu}^\text{(V)} \in \mathbb{R}^{3N_c}$ and $M \in \mathbb{R}^{3N_c \times 3N_c}$. 
Solving Eq.~(\ref{eq_mu^n}) or its matrix form is not feasible with direct method and iterative methods like SOR can be used \cite{Schnieders:2007}. {\bf Note:} $M$ is symmetric and with 1 on diagonal and there might be potentially more convenient numerical methods to solve for $\boldsymbol{\mu}^\text{(V)}$ considering the special structure of $M$ (to be investigated). 
The Coulomb potential $G^\text{V}$ due to $\boldsymbol{\mu}^\text{(V)}$ can be expressed in form of Green's functions 
\begin{equation}\label{eq_GV}
G^\text{V}({\bf r}) = \sum_{n=1}^{N_c}  \frac{r_i - r_{n,i}}{|{\bf r}-{\bf r}_n|^3} \mu^\text{(V),n}_i, \quad
E^\text{(V)}_\text{elec} = \frac{1}{2} k_B T \int \left(G^{\bf M}({\bf r})+G^\text{V}({\bf r})\right) \sum_{n=1}^{N_c} \rho^n({\bf r}) \, d{\bf r}
\end{equation}
After determining the Coulomb potential for permanent multipoles and induced dipoles, i.e., $G^{\bf M}$ and $G^\text{V}$, the electrostatic energy in vacuum $E^\text{(V)}_\text{elec}$ can then be calculated according to the definition in (\ref{eq_GV}).

\subsubsection{Polarization in the solvated phase}

The electrostatic interaction between the solute and the solvent has to be taken into consideration when we study polarization in solvated phase. 
Consider a macromolecule with a low dielectric $\varepsilon^-$ immersed in a solvent with a high dielectric $\varepsilon^+$. The 
equilibrated state of the self-consistent process in the solvent can {\it only} be determined iteratively, because the polarization is now not only {\it non-additive}, but also {\it inseparable}. We propose to characterize this polarization with three components 
as shown in Fig.~\ref{fig:domains}(b):\\
(a) Direct induction by other permanent multipoles; \\
(b) Mutual induction by other induced dipoles; \\
(c) Polarization induced by the total reaction field $\phi_\text{RF}$ as shown at the bottom, 
\begin{equation}\label{eq_mun2}
\boldsymbol{\mu}^n = \alpha^n  \left( \sum_{m \ne n}  \nabla G^m({\bf r}_n) 
+ \sum_{m \ne n}  {\bf T}_{nm} \boldsymbol{\mu}^m 
- \nabla \phi_\text{RF}({\bf r}_n) \right),
\end{equation}
where the reaction 
potential $\phi_\text{RF}$ is the difference between  the electrostatic potential $\phi$ and Coulomb potential $G$, i.e., $\phi_\text{RF} = \phi - G$. Here the electrostatic potential $\phi$ satisfies the PB equation  (\ref{LPBE})  with the total singular source given by
\begin{equation}\label{eq_rho}
\rho = 4 \pi  \sum_{n=1}^{N_c}\left(q^n\delta({\bf r}-{\bf r}_n)+(\mu_i^n+d_i^n)\partial_i\delta({\bf r}-{\bf r}_n)
+Q^n_{ij}\partial_{ij}\delta({\bf r}-{\bf r}_n)\right).
\end{equation}
Unlike Eq.~(\ref{eq_qn}), in which only the permanent multipoles are involved, $\rho$ defined in Eq. (\ref{eq_rho}) contains both permanent multipoles and induced dipoles. In particular, at ${\bf r}_n$ we have the total dipole ${\bf p}^n = {\bf d}^n + \boldsymbol{\mu}^n$. Note although quadrupoles can be polarized, we omit this consideration since quadrupole  polarization responds to the field gradient is weaker compared with dipole's respondence to the more stronger and prevalent electric field. 
In terms of energies, the dipole-field interaction energy scales as $E_\text{dipole} \approx \vec{p} \cdot \vec{E}$ while the quadrupole-filed interaction energy scales as $E_\text{equd} \approx \vec{Q} : \nabla{\vec{E}}$, thus dipoles interact at lower order and with larger energy magnitude, making their effects more significant. Furthermore, consider the electricity density distribution, dipole induction just requires shifting charge clouds slightly,
while quadrupole induction requires differential distortion, which is more complex.\\


In summary, the proposed PMPB system consists of Eqs. (\ref{LPBE}), (\ref{eq_mun2}) and (\ref{eq_rho}) to govern the recursive self-consistent mutual polarization process. Briefly, it consists of the following four steps in each cycle: \\
(i). Calculate $\rho$ by Eq. (\ref{eq_rho}) based on $\boldsymbol{\mu}$ (if for the first time, use some initial guess of $\boldsymbol{\mu}$);\\
(ii). Solve Eq. (\ref{LPBE}) for $\phi$ with the source $\rho$; \\
(iii). Find reaction potential $\phi_\text{RF} = \phi - G$;\\
(iv). Calculate $\boldsymbol{\mu}$ by Eq. (\ref{eq_mun2}) based on $\phi_\text{RF}$. 

In the PMPB models, 
the governing equation of the equilibrated state solution $\boldsymbol{\mu}$ 
of the self-consistent mutual polarization is still linear, 
so that theoretically one can directly invert a linear operator for $\boldsymbol{\mu}$. 
However, with the same consideration as shown in the vacuum phase, it is more convenient to address the self-consistent mutual polarization in (iv) iteratively using say SOR, 
repeating the cycle $\boldsymbol{\mu}\Rightarrow \rho \Rightarrow \phi \Rightarrow \phi_\text{RF} \Rightarrow \boldsymbol{\mu}$, until the steady state. We next provide some details for solving the ${\bf \mu}$ from Eq.~(\ref{eq_mun2}). 

{\bf Self-consistent Polarization using SOR.} 


\subsubsection{Calculation of electrostatic free energy for biological applications}
After studying polarizations in both vacuum and solvent, the electrostatic free energy $E_\text{elec}$ can be calculated based on the following physical considerations. 
Let us abuse the notation and denote the induced dipole and potential in the equilibrium state as $\boldsymbol{\mu}$ and  $\phi$, respectively. A three-component decomposition is first conducted: $\phi = \phi_\text{RF} + G^{\bf M} + G^{\boldsymbol{\mu}}$, where $G^{\boldsymbol{\mu}}$ can be similarly defined as $G^\text{V}$ in Eq.~(\ref{eq_GV}). Likewise, the electrostatic energy in solvent $E^\text{(S)}_\text{elec}$ can be defined 
by replacing  $G^{\bf M}({\bf r})$ and $G^\text{V}({\bf r})$ by $\phi$ in  Eq.~(\ref{eq_GV}). The electrostatic free energy $E_\text{elec}$ measuring the difference between $E^\text{(S)}_\text{elec}$ and $E^\text{(V)}_\text{elec}$ then consists of two terms, i.e., the reaction field $\phi_\text{RF}$ and the difference between the self-consistent induced dipoles in solvent and vacuum $G^{\Delta} = G^{\boldsymbol{\mu}} -  G^\text{V}$. Thus, we propose to calculate $E_\text{elec}$ as
\begin{equation}\label{eq_E}
E_\text{elec}=\frac{1}{2}k_B T \sum_{n=1}^{N_c} q^n [\phi_\text{RF}({\bf r}_n) + G^{\Delta}({\bf r}_n) ]
+ d^n_i \partial_i [\phi_\text{RF}({\bf r}_n) + G^{\Delta}({\bf r}_n) ]
+ Q^n_{ij} \partial_{ij} [\phi_\text{RF}({\bf r}_n) + G^{\Delta}({\bf r}_n) ]. 
\end{equation}
Note that with the equilibrated $\boldsymbol{\mu}$ and $\boldsymbol{\mu}^\text{(V)}$, 
$G^{\Delta}$ and its derivatives can be calculated analytically in Eq.~(\ref{eq_E}). Excluding self-interacting energy, we have  
$ G^{\Delta}({\bf r}_n) = \sum_{m=1,m \ne n}^{N_c} \left(\mu^m_i - \mu^{\text{(V)},m}_i \right) \displaystyle\frac{r_{n,i} - r_{m,i}}{|{\bf r}_n-{\bf r}_m|^3}$.


\subsubsection{Regularization formulation}
In regularizing the singular source in Eq.~(\ref{LPBE}), we propose to solve the PM-NPB equation by a two-component decomposition  $\phi = \phi_\text{RF}+\phi_C$ \cite{Geng:2017, Lee:2020}, 
where the Coulomb potential $\phi_C$ given in terms of Green's function as
\begin{equation}\label{eq_phiC}
\phi_C({\bf r})=G({\bf r}) :=
C \sum\limits_{n=1}^{N_c}\frac{1}{\varepsilon^- } 
\left[\frac{1}{|{\bf r}-{\bf r}_n|}q^n+\frac{r_i-r_{n,i}}{|{\bf r}-{\bf r}_n|^3}p^n_i+\frac{(r_i-r_{n,i})(r_j-r_{n,j})}{2|{\bf r}-{\bf r}_n|^5}Q^n_{ij} \right]
\end{equation}
solves  the Gauss's law  in the free space
\begin{equation} \label{phi_C}
- \varepsilon^- \Delta \phi_C({\bf r})=4 \pi C \displaystyle\sum\limits_{n=1}^{N_c}q^n\delta({\bf r}-{\bf r}_n)+p^n_i\partial_i\delta({\bf r}-{\bf r}_n)
+Q^n_{ij}\partial_{ij}\delta({\bf r}-{\bf r}_n).
\end{equation}
By capturing the singularities via $\phi_C$, the reaction field potential $\phi_\text{RF}$ satisfies
\begin{eqnarray}
- \varepsilon^- \Delta \phi_\text{RF} = & 0, & \text{ in } \Omega^- \label{eq_npb1}\\
- \varepsilon^+ \Delta \phi_\text{RF} + \bar{\kappa}^2 \sinh (\phi_\text{RF} + G) = & \varepsilon^+ \Delta G, & \text{ in } \Omega^+  \label{eq_npb2}\\
\left[\phi_\text{RF}\right]=0, \quad 
\left[ \varepsilon\frac{\partial \phi_\text{RF}}{\partial \nu} \right]= & (\varepsilon^- - \varepsilon^+) \frac{\partial G}{\partial \nu},
&\text{ on } \Gamma  \label{eq_npb3} \\
\phi_\text{RF} =& \phi_b - G, &\text{ on } \partial\Omega  \label{eq_npb4}
\end{eqnarray}
where the derivative of $G$ is known analytically.

\section{Numerical Results}
\subsection{AMOEBA Forcefield Inclusion}

Users will need to download the Tinker software package first.
This can be done using command line \texttt{git clone https://github.com/TinkerTools/Tinker}. 
Then, a Brookhaven PDB file downloaded from Protein Data Bank (https://www.rcsb.org/) can be converted into a Tinker \texttt{.xyz} Cartesian coordinate file using the program \texttt{pdbxyz} in Tinker. This conversion can be done under \texttt{source} subdirectory under Tinker directory using the command line: \texttt{pdbxyz.x PDBID.pdb -key PDBID.key}.
Here, the key file is a single line file, i.e., ``parameters amoebapro13", which can redirect to the force field file \texttt{amoebapro13.prm}. More parameter files can be found under \texttt{params} subdirectory under Tinker directory.\\

Next, a \texttt{.xyz} file needs to be converted to a dummy \texttt{.pqr} file which contains charge positions, point charges, dipole moments, and quadrupole moments information, plus a \texttt{.xyzr} file which contains the charge position and radius information. The \texttt{.xyzr} file will also be used when calling \texttt{MSMS} molecular surface software. 
The \texttt{.pqr} file can be generated using the updated python script under \texttt{src/test\_proteins} under MIB directory by the command line \texttt{python readData.py PDBID.xyz}. The \texttt{readData.py} is originally from the software PYGBE (https://github.com/cdcooper84/pygbe).
The \texttt{.xyzr} file can be generated using the python script under \texttt{src/test\_proteins} under MIB directory by the command line \texttt{python tinker\_to\_xyzr.py PDBID}. This file is also from software PYGBE (https://github.com/cdcooper84/pygbe).

\subsection{Spherical Results}

Our numerical results 
are computed using a 13inch MacBook Pro with intel core-i5 processor and 16 GB of RAM.
The dielectric constants of the solvent domain and molecular domain for the Kirkwood's results are set as 80 and 1 respectively. For the test proteins in Table \ref{PMPBresults}, the solvent dielectric constant is set as 78.3, and $\kappa$ is set as 0.125 \AA$^{-1}$.
Note that we only use permanent dipole moments
discussed in Section \ref{sec:solvatephase}. 

As shown in Table \ref{Kirkwoodresults}, dcel is the grid size for finite difference method, $e_{\text{int}}$ is the interface error, Column 3 is the corresponding order of convergence. $E_{\text{sol}}$ is the solvation energy (kcal/mol), $e_{E_{\text{sol}}}$ is the corresponding error with Column 6 as the order of convergence.
These tables show a $2^{\text{nd}}$ order of convergence, validating the fitness of PM source with our MIB-PB solver. Comparing the values of $E_{\text{sol}}$ for monopole and multipole in Column 5, it shows that using the multipole moments significantly enhances the modeling accuracy. Note these results are consistent with the test cases reported in \cite{Schnieders:2007}

\begin{table}[htbp]
\caption{Kirkwood's spherical cavity results with radius $2$\AA.}
\centering
\begin{tabular}{l||l||l||l||l||l}
\hline\hline
\multicolumn{6}{c}{Centered monopole in a spherical cavity} \\ \hline \hline
dcel & $e_{\text{int}}$ & order & $E_{\text{sol}}$   & $e_{E_{\text{sol}}}$ & order \\ \hline
1   &   4.46E-04    &      & -81.9611 & 1.90E-02  &  \\ \hline
0.5 & 7.31E-05 & 2.61  & -81.9718 & 8.35E-03  & 1.19  \\ \hline
0.25  & 2.07E-05  & 1.82  & -81.9784 & 1.73E-03   & 2.27  \\ \hline
0.125 & 5.67E-06  & 1.87  & -81.9798 & 3.38E-04   & 2.36  \\ \hline
0.0625 & 1.40E-06    & 2.02  & -81.9801 & 6.30E-05    & 2.42  \\ \hline \hline                
\multicolumn{6}{c}{Centered dipole in a spherical cavity} \\ \hline \hline
1 & 2.86E-04    &       & -2.3939 & 2.27E-03      &                           \\ \hline
0.5 & 5.85E-05    & 2.29  & -2.3949 & 1.27E-03                       & 0.84                      \\ \hline
0.25& 1.82E-05    & 1.68  & -2.3959 & 2.81E-04                       & 2.18                      \\ \hline
0.125   & 4.91E-06    & 1.89  & -2.3962 & 5.41E-05                       & 2.38                      \\ \hline
0.0625 & 1.22E-06    & 2.00  & -2.3962 & 9.95E-06                       & 2.44                      \\ \hline     \hline          
\multicolumn{6}{c}{Centered quadrupole in a spherical cavity} \\ \hline \hline
1& 3.78E-04    &       & -1.7883 & 4.17E-03                       &                           \\ \hline 
0.5& 1.05E-04    & 1.85  & -1.7901 & 2.38E-03                       & 0.81                      \\ \hline 
0.25& 3.67E-05    & 1.52  & -1.7919 & 5.37E-04                       & 2.15                      \\ \hline 
0.125& 1.07E-05    & 1.77  & -1.7923 & 1.01E-04                       & 2.41                      \\ \hline 
0.0625& 2.65E-06    & 2.02  & -1.7924 & 1.82E-05                       & 2.47                      \\ \hline \hline
\multicolumn{6}{c}{Centered multipole in a spherical cavity} \\ \hline \hline
1& 9.79E-04 &       & -86.1428                     & 2.60E-02             & \multicolumn{1}{l}{} \\
0.5 & 1.88E-04 & 2.38  & -86.1565                     & 1.24E-02             & 1.07                 \\
0.25& 5.77E-05 & 1.70  & -86.1662                     & 2.62E-03             & 2.24                 \\
0.125& 1.56E-05 & 1.88  & -86.1683 & 5.04E-04             & 2.38                 \\
0.0625 & 3.67E-06 & 2.09  & -86.1687                     & 9.33E-05             & 2.43                 \\ \hline \hline
\end{tabular}
\label{Kirkwoodresults}
\end{table}

As outlined in Section \ref{appendix:BC}, the boundary conditions of our PM-PB model rely on selecting the radius \(a\) for the approximated ``sphere". \textbf{Further} investigation is needed to fully understand this concept.
In Table \ref{PMPBresults}, we use $a=60$\AA~ to compute the results for several test proteins. We compute the solvation energy value at \(1/\infty\) by linearly extrapolating it as the grid size approaches infinity. We consider this value as our exact value and compare the results with it using the formula:
\begin{equation}
\text{Error} = \frac{\left|E_{\text {sol }}-E_{\text {sol }}^{ex}\right|}{\left|E_{\text {sol }}^{ex}\right|} \times 100\%.
\end{equation}
The results for these proteins show a convergent pattern validating the performance of our MIB-PB solver coupled with PM source term.

\begin{table}[htbp]
\caption{Results on test proteins with atom number ranging from $504-1046$; showing electrostatic solvation energy $E_{\text{sol}}$ (kcal/mol) where the value for $\frac{1}{\infty}$ is linearly extrapolated. The error is computed based on this extrapolated value.
}
\centering
\resizebox{\textwidth}{!}{
\begin{tabular}{l||ll||ll||ll||ll||ll}
\hline\hline
      & \multicolumn{2}{c||}{1crn}     & \multicolumn{2}{c||}{1enh}     & \multicolumn{2}{c||}{1fsv}     & \multicolumn{2}{c||}{1pgb}     & \multicolumn{2}{c}{1vii}     \\ \hline
h     & $E_{\text{sol}}$& Error & $E_{\text{sol}}$ & Error  & $E_{\text{sol}}$ & Error& $E_{\text{sol}}$ & Error & $E_{\text{sol}}$ & Error \\
1.00  & -230.51         & 0.68\%     & -1476.61        & 1.66\%     & -777.67         & 1.64\%     & -797.66         & 1.00\%     & -681.62         & 0.49\%     \\
0.50  & -227.98         & 0.43\%     & -1450.01        & 0.17\%     & -763.14         & 0.25\%     & -785.22         & 0.57\%     & -676.64         & 0.24\%     \\
0.25  & -228.47         & 0.21\%     & -1451.28        & 0.08\%     & -764.12         & 0.13\%     & -787.49         & 0.29\%     & -677.47         & 0.12\%     \\
$\frac{1}{\infty}$ & -228.96         &            & -1452.50        &            & -765.09         &            & -789.76         &            & -678.29         &           \\ \hline\hline
\end{tabular}}
\label{PMPBresults}
\end{table}


\section {Software Dissemination}

\section {Concluding Remarks}

\section{Appendix}
\subsection{Analytic Solutions for test cases using Kirkwood sphere}

Kirkwood's dielectric sphere is widely recognized as a robust benchmark for assessing the effectiveness of Poisson-Boltzmann (PB) solvers in terms of accuracy, convergence speed, and efficiency. Within this framework, we explore the analytical solutions derived from 
a point monopole, 
a point dipole, 
and a point quadrupole, each positioned at the center of a spherical cavity respectively.
In addition, 
we consider a centered induced dipole representing the polarization of the solute. 

\subsubsection{centered monopole in a spherical cavity}

For a centered monopole in a dielectric sphere. The analytical solution has a closed form 
as in
\begin{subequations}
\begin{align}
\phi_{\text {out }} & =
\frac{q}{ r}, \\
\phi_{\text {in}} & =
\left[\frac{1}{ r}-\frac{\epsilon_2-\epsilon_1}{\epsilon_2} \frac{1}{ a}\right] q,
\end{align}
\end{subequations}
where $r$ is the distance between the potential measured and the centered charge $q$.
The potential energy, with subscript $m$ denoted for monopole, is:
\begin{equation}
U_m=-\frac{1}{2}\left(\frac{\epsilon_2-\epsilon_1}{\epsilon_2}\right) \frac{q^2}{a}.
\end{equation}

\subsubsection{centered dipole in a spherical cavity}
Similarly, for a point dipole, we derive:
\begin{subequations}
\begin{align}
\phi_{\text{out}} &=\frac{3 \epsilon_2}{2 \epsilon_2+ \epsilon_1} \frac{1}{r^3} \mathbf{d} \cdot \mathbf{r}, \\
\phi_{\text{in}} &=\left[\frac{1}{r^3}-\frac{2(\epsilon_2-\epsilon_1)}{2 \epsilon_2+\epsilon_1} \frac{1}{a^3}\right] \mathbf{d} \cdot \mathbf{r}.
\end{align}
\end{subequations}
where 
$\mathbf{d}$ stands for the dipole moment vector and $\mathbf{r}$ is the direction vector.
The potential energy, with subscript $d$ denoted for dipole, is:
\begin{equation}
U_d=-\frac{1}{2}\left(\frac{2(\epsilon_2-\epsilon_1)}{2 \epsilon_2+\epsilon_1} \frac{1}{a^3} \right) \mathbf{d} \cdot \mathbf{d}.
\end{equation}
\subsubsection{centered quadruple in a spherical cavity}
Finally, for a point quadrupole, we derive:
\begin{subequations}
\begin{align}
\phi_{\text{out}} &=\frac{5 \epsilon_2}{3 \epsilon_2 + 2 \epsilon_1} \frac{3}{r^5} \Theta : \mathbf{r} \mathbf{r}, \\
\phi_{\text{in}} &=\left[\frac{1}{r^5}-\frac{3\left(\epsilon_2-\epsilon_1\right)}{3 \epsilon_2+2\epsilon_1} \frac{1}{a^5}\right] 3 \Theta : \mathbf{r} \mathbf{r},
\end{align}
\end{subequations}
where $\Theta$ the quadrupole moment and $\mathbf{r} \mathbf{r}$ are 3-by-3 tensors. Their product $\Theta : \mathbf{r} \mathbf{r}$ is a scaler, performing the same way as a dot product.
As the traceless quadrupole is normally used,
the potential derived needs to be shifted by a coefficient $\frac{1}{3}$:
\begin{subequations}
\begin{align}
\phi_{\text{out}} &=\frac{5 \epsilon_2}{3 \epsilon_2 + 2 \epsilon_1} \frac{1}{r^5} \Theta : \mathbf{r} \mathbf{r}, \\
\phi_{\text{in}} &=\left[\frac{1}{r^5}-\frac{3\left(\epsilon_2-\epsilon_1\right)}{3 \epsilon_2+2\epsilon_1} \frac{1}{a^5}\right] \Theta : \mathbf{r} \mathbf{r}.
\end{align}
\end{subequations}
The potential energy, with subscript $q$ denoted for quadrupole, also shifted by a coefficient $\frac{1}{3}$, is represented as:
\begin{equation}
U_q=-\frac{1}{6}\left(\frac{3\left(\epsilon_2-\epsilon_1\right)}{3 \epsilon_2+2\epsilon_1} \frac{1}{a^5} \right) \Theta \Theta.
\end{equation}
\subsubsection{centered multipole in a spherical cavity}

\noindent
{\bf Acknowledgments} 

This work was supported in part by the National Science Foundation (NSF) grant DMS-2110922 and DMS-2110869. 

We thank SMU Math Department by providing Parallel Computing Class Math 6370, which systematically trained graduate students with parallelization strategies, schemes, and experience. We Also thank SMU Center for Research Computing (CRC) by proving computing hardware. These resources combined make this project possible.

\bibliographystyle{acm}
\bibliography{pb_review_um}

\end{document}